# Numerical simulation of oscillating topological solitons in 2D O(3) nonlinear sigma model

F. Sh. Shokirov

S. U. Umarov Physical-Technical Institute of Academy of Sciences of the Republic of Tajikistan, Aini Avenue 299/1, Dushanbe, 734063

E-mail: farhod0475@gmail.com

**Abstract:** Dynamics of interaction of topological solitons (vortices) in (2+1)-dimensional O(3) nonlinear sigma model in anisotropic case are investigated. By numerical simulation methods is shown that the changes of rotation frequency of isotopic spin in the fiber space leads to an oscillatory dynamics of topological solitons. The models of interaction of oscillating topological solitons are obtained and their properties are investigated.

## I. Introduction

Nonlinear nonperturbative models of field theory, beginning with the pioneering works of T.H.R. Skyrme [1] are attracted unrelenting attention. In nonperturbative models, unlike the Higgs model, the elementary particles find their natural description as "clumps" of energy. Another important advantage of these models is the absence of divergences, which in the standard model require the use of renormalizations techniques. At the same time, as a model representation of the Yang-Mills equations in the field theory is also offered class of SU(N) two-dimensional nonlinear sigma models [2].

In this paper is conducted numerical studies of nonlinear excitations with nontrivial Hopf index in (2+1)-dimensional (2D) anisotropic O(3) nonlinear sigma model (NLSM) with particular attention to the dynamics of the interaction of oscillating topological objects. At the first stage, we obtained numerical models of topological solitons (TS, topological vortices) of Belavin-Polyakov type [3], having the dynamics of internal degrees of freedom (oscillating TS). Recall that the Euler-Lagrange equation of 2D O(3) NLSM [4-11] have the following form:

$$\partial_\mu \partial^\mu s_i + \left(\partial_\mu s_a \partial^\mu s_a\right)s_i - s_3(\delta_{i3} - s_i s_3) = 0,$$
$$i = 1, 2, 3, \quad s_a s_a = 1,$$
(1)

where $s_1 = \sin\theta\cos\varphi$, $s_2 = \sin\theta\sin\varphi$, $s_3 = \cos\theta$, the $\theta(x,y,t)$ and $\varphi(x,y,t)$ are the Euler angles.

In present paper is conducted numerical study of the TS of the form

$$\theta = 2arctg\left(\frac{r}{R}\right)^{Q_t}, \quad \varphi = Q_t\chi - \omega\tau, \tag{2}$$
$$r^2 = x^2 + y^2, \; R = 1, \; \cos\chi = \frac{x}{r}, \; \sin\chi = \frac{y}{r},$$

where $Q_t = 3$ – topological charge (Hopf index) for values of frequency of rotation A3-field vector – $\omega = \left\{\frac{1}{2}, 1, 2\right\}$ in the isospace of sphere $S^2$.

Dynamics of interaction of TS (2) of the model (1) in the case of $\omega = 1$ for $Q_t \in \{1,2,...,6\}$ were investigated in our previous works (see, e.g., [5-10]), where were discovered a number of properties, including:

– TS decay onto the localized perturbations (a preserving the $Q_t$ sum);
– a phased annihilation of TS by the units of $Q_t$ (by radiation of energy in the form of linear perturbations waves propagating with speed of $c$);
– long-range interaction of TS;
– mutual attraction and repulsion of TS, etc.

We note that in our previous works models of interaction of TS (2) with a 180-degree domain wall of Neel type (see, e.g., [9,10]) were also investigated. It was shown that the TS (2) at the collision with the domain wall is completely decay by a phased radiation of energy in the form of $2Q_t$ localized perturbations (LP), each of which has $|Q_t| = \frac{1}{2}$ and moves along the domain wall with a speed $v \to c$ ($c = 1$ - speed of light). The energy integral ($En$) of interacting TS system were preserved with accuracy: $\frac{\Delta En}{En} \approx 10^{-6} - 10^{-3}$.

In the second part of this paper presents the results of numerical investigation of stationary TS with the different values ($\omega \neq 1$), where are obtained a numerical models of oscillating TS.

The third part is devoted to the study of the TS interactions dynamics, where in particular describes a models of head-on collisions and TS decays onto the LP. In the final part of the paper summarizes results and put forward certain proposals and conclusions.

## II. The stationary oscillating topological solitons

In this part of the paper presents the results of a numerical study of the TS (2) of model (1) at $\omega = \frac{1}{2}$ and $\omega = 2$.

**Case $\omega = \frac{1}{2}$.** In Fig.1 shows the process of evolution (of energy density: $DH$) of TS (2) at $\omega = \frac{1}{2}$; time simulation $t \in [0.0, 120.0]$. The TS energy density is concentrated in a radially symmetrical annular structure [4-10]. Throughout the simulation time ($[0.0, 120.0]$) occurs the 12 uniform oscillation periods of TS. The first period is shown in Fig.1a - Fig.1c. In Fig.1d shown final state of the vortex evolution ($t = 120$). The energy integral of evolving TS after the formation of oscillating soliton ($t > 40.0$) is preserved with high accuracy: $\frac{\Delta En}{En} \approx 10^{-6}$ (Fig.1e).

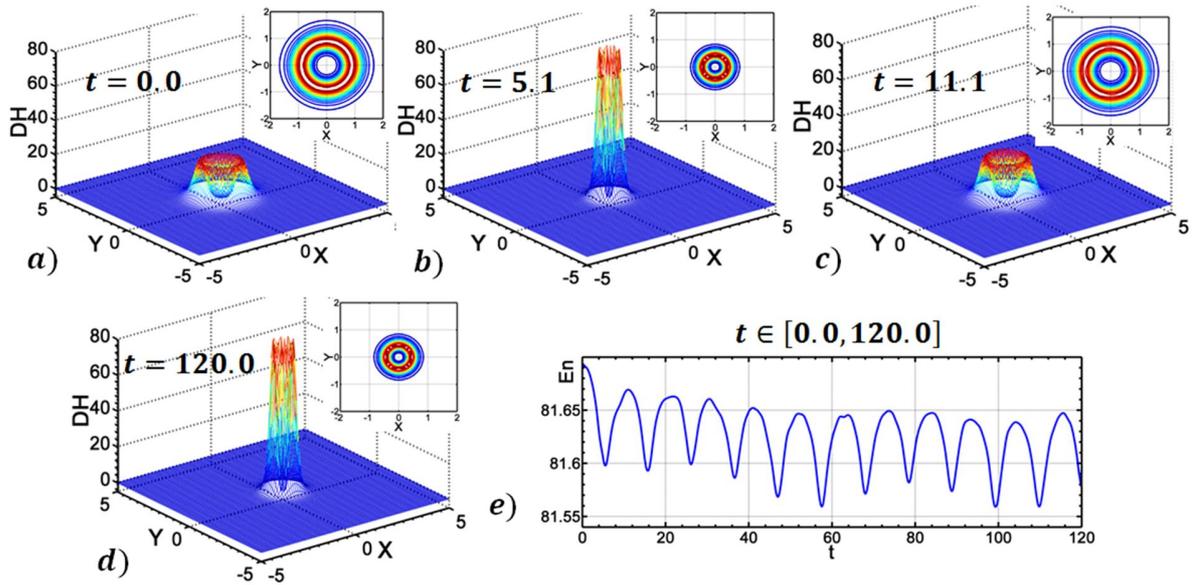

**Fig.1.** Evolution ($DH$ and its contour projection) of stationary ($v = 0.0$) TS (2) of model (1) with the Hopf index $Q_t = 3$ at $\omega = 0.5$: **a)** $t = 0.0$; **b)** $t = 5.1$; **c)** $t = 11.1$; **d)** $t = 120.0$; **e)** energy integral for $t \in [0.0, 120.0]$.

The feature of dynamics of oscillating TS (2), shown in Fig.1 consist is that the oscillating process occurs only on the contour of ring-shape of energy density concentration ($DH$) (Fig.2a). In this time the center of TS − $DH$ (0,0) remains practically is motionless: $\Delta DH_0 < 2.1e^{-3}$ (Fig.2b).

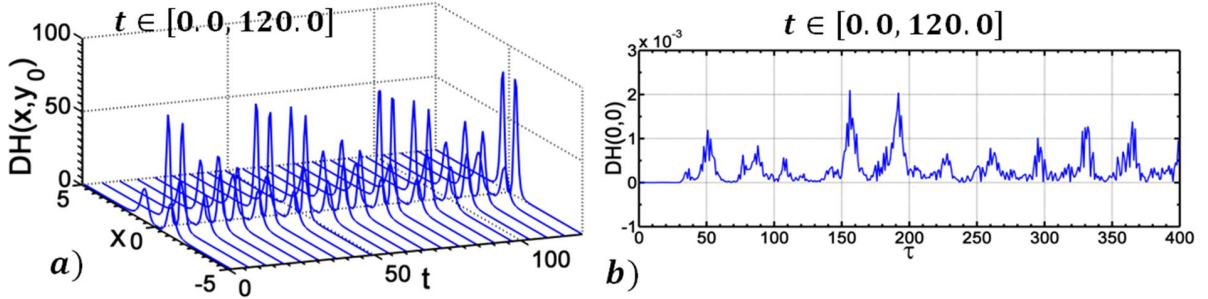

**Fig.2.** Evolution of ($DH$) stationary ($v = 0.0$) TS (2) of model (1) with the Hopf index $Q_t = 3$ at $\omega = 0.5$: **a)** in a planar section - $y = 0$; **b)** dynamics of TS center point - $DH(0,0)$. Time simulation: $t \in [0.0, 120.0]$, ($t \approx 0.3\tau$).

In Fig.3 shown the isospin projection of TS (2) of model (1) on the phased plane $z(x, y)$. These projections have are characteristic form [5-10], but in this case ($\omega = \frac{1}{2}$) are different from of the similar projections of TS (2), which were studied at $\omega = 1$, by periodic change of the annular structure of $DH$ in which concentrated the energy of vortex.

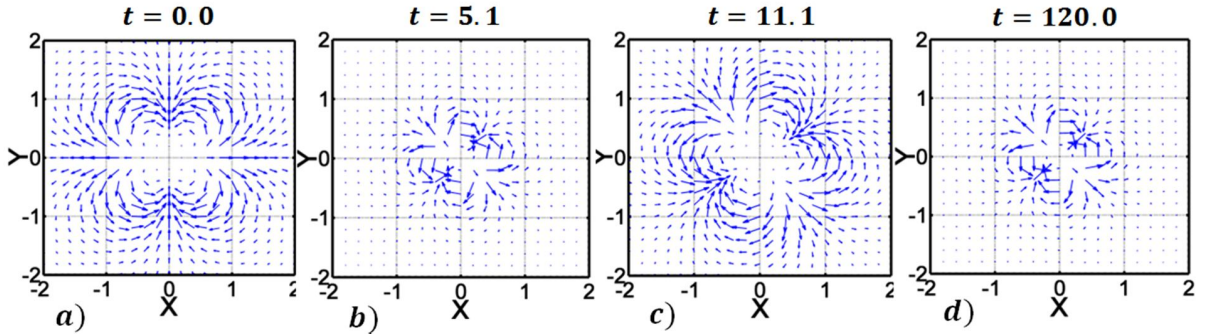

**Fig.3.** Isospin projection of stationary ($v = 0.0$) TS (2) of model (1) with the Hopf index $Q_t = 3$ onto the phase plane $z(x, y)$ at $\omega = 0.5$: **a)** $t = 0.0$; **b)** $t = 5.1$; **c)** $t = 11.1$; **d)** $t = 120.0$.

**Case $\omega = 2$.** In Fig.4 shown the process of evolution (of energy density: $DH$) of TS (2) at $\omega = 2$; time simulation $t \in [0.0, 120.0]$. The TS energy density in this case also is concentrated in a radially symmetrical annular structure [4-10]. Throughout the simulation time ([0.0, 120.0]) occurs the 10 uniform oscillation periods of TS. But in this ($\omega = 2$) case, the character of the oscillations is different from the previous experiment. In the case of $\omega = \frac{1}{2}$ at $t = 0.0$ the energy density ($DH$) has the minimum value (Fig.1a), and this condition varies periodically (at $t = 12.3$ in Fig.1c etc.) with the maximum values (see for example, Fig.1b at $t = 5.1$; the radius of the ring-shaped structure of $DH$ is reduced). In the case of $\omega = 2$ the energy density ($DH$) of TS at $t = 0.0$ (also at $t = 12.3$, etc.) has the maximum values (Fig.4ac). Between these states the energy density ($DH$) of TS has a minimum values (see, for example, Fig.4b at $t = 5.1$; the radius of the ring-shaped structure of $DH$ increases). In Fig.4d shown final state of the vortex evolution ($t = 120$). The energy integral of evolving TS after a formation of oscillating soliton ($t > 40.0$) is preserved with accuracy: $\frac{\Delta En}{En} \approx 10^{-3}$ (Fig.4e). Note that on the simulation area edges are set absorbing boundary conditions, and in this case, increase of radius of the $DH$ ring-shaped structure leads to a relatively large energy dissipation of TS.

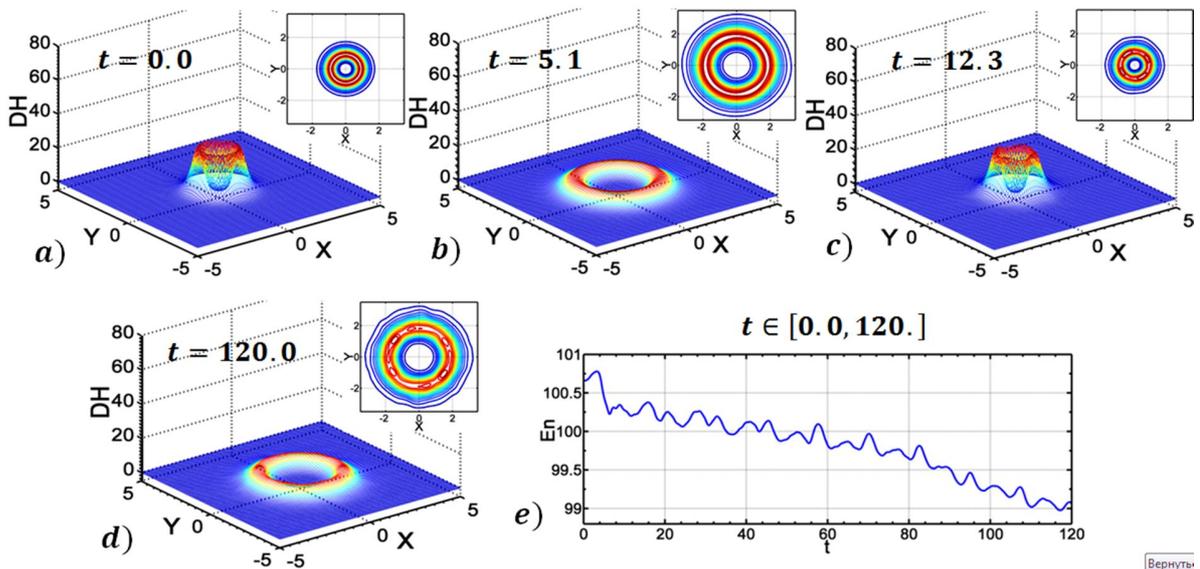

**Fig.4.** Evolution ($DH$ and its contour projection) of stationary ($v = 0.0$) TS (2) of model (1) with the Hopf index $Q_t = 3$ at $\omega = 2.0$: **a)** $t = 0.0$; **b)** $t = 5.1$; **c)** $t = 12.3$; **d)** $t = 120.0$; **e)** energy integral for $t \in [0.0, 120.0]$.

In Fig.5 shown the evolution of TS, that is described in Fig.4 in a planar section $y = 0$ at $t \in [0.0, 120.0]$. Similarly TS with $\omega = \frac{1}{2}$ in these case oscillations is observed only on the contour of the ring-shape structure of the TS energy density (Fig.5a). In this time the center of TS – $DH$ (0,0) remains practically is motionless: $\Delta DH_0 < 2.4e^{-2}$ (Fig.5b).

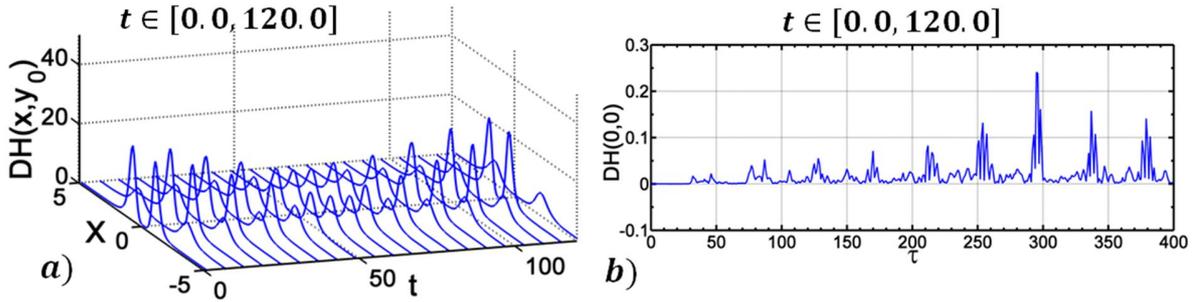

**Fig.5.** Evolution of ($DH$) stationary ($v = 0.0$) TS (2) of model (1) with the Hopf index $Q_t = 3$ at $\omega = 2.0$: **a)** in a planar section $y = 0$; **b)** dynamics of TS center point: $DH(0,0)$. Time simulation: $t \in [0.0, 120.0]$, ($t \approx 0.3\tau$).

In Fig.6 shown the isospin projection of TS (2) of model (1) on the phased plane $z(x, y)$. These projections have are characteristic form [5-10], but and in this case ($\omega = 2.0$) are different from of the similar projections of TS (2), which were studied at $\omega = 1$, by the periodic change of the annular structure of the $DH$ in which concentrated the energy of the vortex.

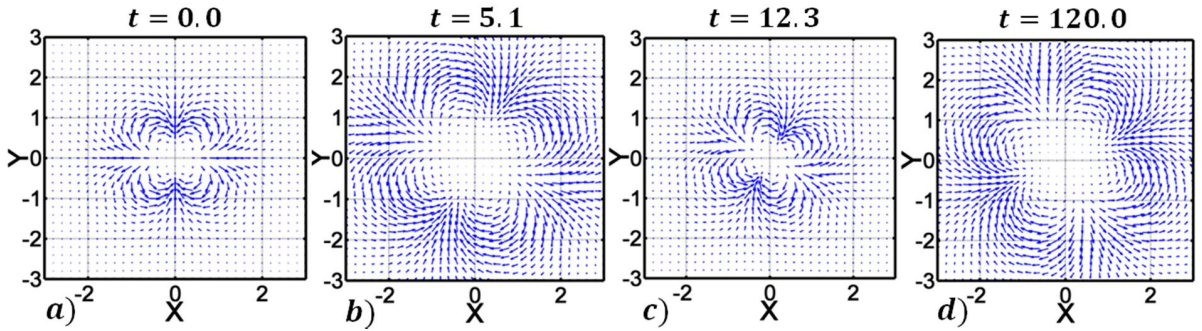

**Fig.6.** Isospin projection of stationary ($v = 0.0$) TS (2) of model (1) with the Hopf index $Q_t = 3$ on the phase plane $z(x, y)$ at $\omega = 2.0$: **a)** $t = 0.0$; **b)** $t = 5.1$; **c)** $t = 12.3$; **d)** $t = 120.0$.

Thus the results of numerical experiments conducted in the second part of the paper showed that in 2D O(3) NLSM the variation of values ($\omega \neq 1$) of frequency of A3-field vector rotation in isospace of sphere $S^2$ leads to a stable oscillating dynamics of the TS.

Note that our numerical models are based on the three-layer finite difference scheme of the second order accuracy $O(h^2 + \tau^2)$ [12,13], with using properties of stereographic projection and taking into account the group-theoretic properties of O(N) NLSM class of the field theory [4-11]. A special algorithm for the difference scheme used in the present paper was developed in work [4] and updated in the works [5-11]. In the next part, based on the developed models of stationary oscillating TS is conducted numerical study of the dynamics of their interaction. Note that the actual dynamics of soliton solutions, where fully can manifest their special, particle-like properties can be obtained carrying out by studies of the dynamics of their interactions [14].

### III. Head-on collisions of oscillating topological solitons

In this part present results of research models of frontal (head-on) collisions of TS (2) in anisotropic 2D O(3) NLSM for different values of frequency of A3-field vector rotation: $\omega = \left\{\frac{1}{2}, 1, 2\right\}$ in the isospace of sphere $S^2$:

- $\omega_L = 1 \rightarrow \leftarrow \omega_R = \frac{1}{2}$;
- $\omega_L = 1 \rightarrow \leftarrow \omega_R = 2$;
- $\omega_L = \frac{1}{2} \rightarrow \leftarrow \omega_R = \frac{1}{2}$;
- $\omega_L = 2 \rightarrow \leftarrow \omega_R = 2$.

It should be noted that the dynamics of the interaction of the TS (2) in the case of

- $\omega_L = 1 \rightarrow \leftarrow \omega_R = 1$,

has been investigated in detail in our previous works (see, e.g., [5-8]).

**Case: $\omega_L = 1 \rightarrow \leftarrow \omega_R = \frac{1}{2}$.** In Fig.7 shown the process of evolution ($DH$ and its contour projection) of frontal collision of TS (2) having a different values of $\omega$. From left moves the TS with $\omega_L = 1$, from right moves the TS with $\omega_R = \frac{1}{2}$ in the opposite directions; time simulation $t \in [0.0, 90.0]$. The initial speed ($t_0 = 0.0$), given by the Lorentz transformation for both TS is: $|v_{LR}(t_0)| \approx 0.0995$. At $t = 37.8$ are both TS moves a distance equal to about $s \approx 3.3$ units (Fig.7b), at $t = 42.3$ (Fig.7c) interact and form a bound a ring-shaped state, in the center of which is focused a dense energy concentration. Next, the TS is separated for a short time (Fig.7d) and again attracted, form a bound a ring-shaped state (Fig.7d) similar to that shown in Fig.7c. The energy integral of the interactions TS system is preserved with accuracy: $\frac{\Delta En}{En} \approx 10^{-4}$ (Fig.7f).

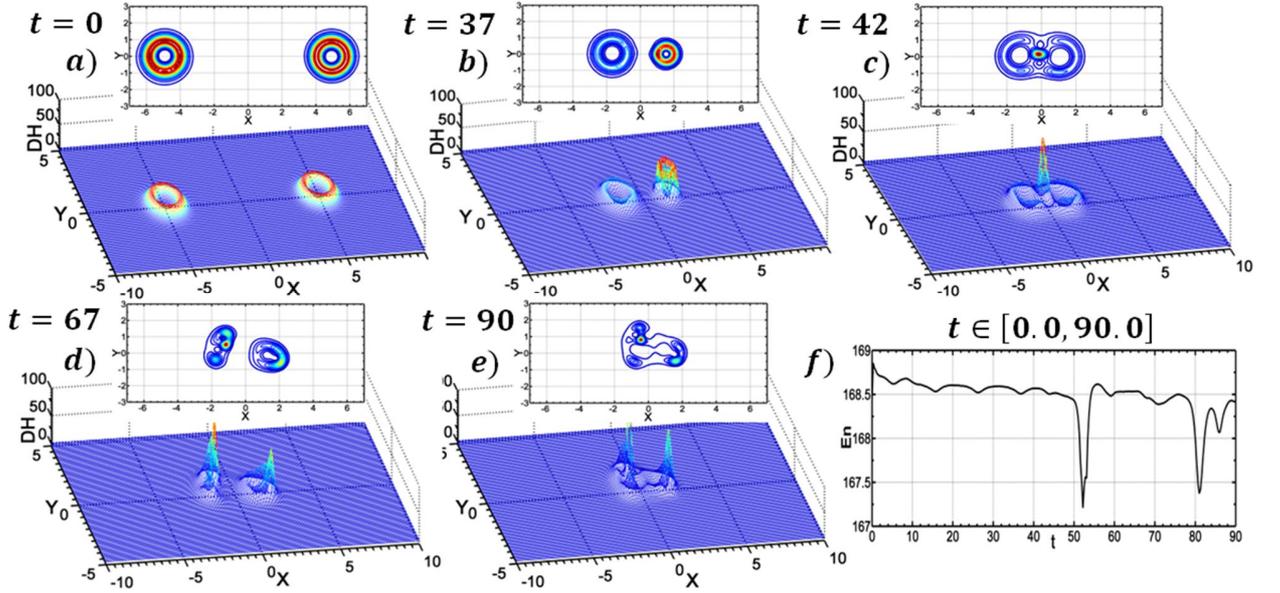

**Fig.7.** Evolution ($DH$ and its contour projection) of head-on collision of TS (2) of model (1) at $Q_{LR} = 3$, $|\vec{v}_{LR}(t_0)| \approx 0.0995$, $\omega_L = 1.0$, $\omega_R = 0.5$: **a)** $t = 0.0$; **b)** $t = 37.8$; **c)** $t = 42.3$; **d)** $t = 67.8$; **e)** $t = 90$; **f)** energy integral for $t \in [0.0, 90.0]$.

Recall that on the simulation area edges are set special boundary conditions absorbing the excess energy emitted by interacting TS, in the form of linear perturbation waves. Below we consider the process of decay of interacting TC (2) onto the localized perturbations (LP).

In Fig.8 shown the process of a TS head-on collision similar to the previous experiment, but in this case, the initial speed (at $t_0 = 0.0$) of TS increased almost twice: $|v_{LR}(t_0)| \approx 0.196$; time simulation $t \in [0.0, 52.0]$.

At $t = 21.9$ the interacting TS form a bound a ring-shaped state (similar to the state shown in Fig.7c), in the center of which focused a dense energy concentration (Fig.8b).

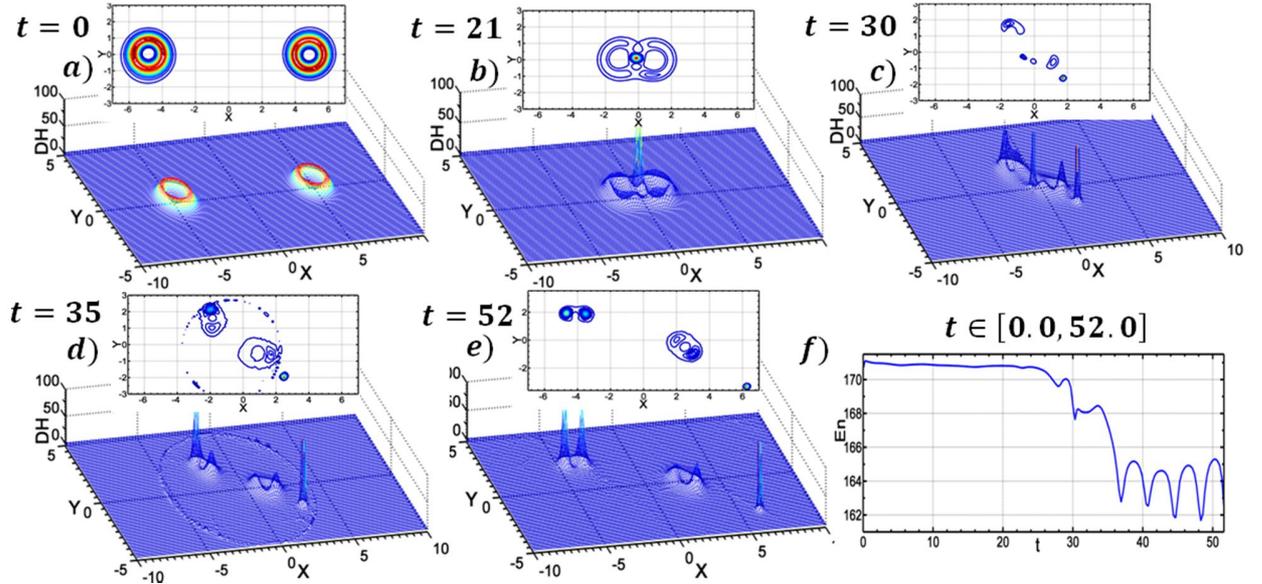

**Fig.8.** Evolution ($DH$ and its contour projection) of head-on collision of TS (2) of model (1) at $Q_{LR} = 3$, $|\vec{v}_{LR}(t_0)| \approx 0.196$, $\omega_L = 1.0$, $\omega_R = 0.5$: **a)** $t = 0.0$; **b)** $t = 21.9$; **c)** $t = 30.0$; **d)** $t = 35.1$; **e)** $t = 52$; **f)** energy integral for $t \in [0.0, 52.0]$.

Next, at $t \approx 30.0$ is observed the TS decay onto the LP with topological charges $Q_t = 1$ (two LP) and $Q_t = 2$ (two LP) (Fig.8c). But after a short time ($t \approx 35.0$) one LP with $Q_t = 1$ is destroyed (annihilation) in form a radially symmetric propagating linear wave (Fig.8d). The rest LP are preserved stability and moving in different directions from the center of the resonance zone (Fig.8e). The energy integral of the interactions TS system after the annihilation of one LP ($t > 35.0$) is preserved with accuracy $\frac{\Delta En}{En} \approx 10^{-4}$ (Fig.8f).

**Case: $\boldsymbol{\omega_R = 1} \rightarrow \leftarrow \boldsymbol{\omega_L = 2}$.** The process of interaction of the TS (2) at $\omega_L = 1$ and $\omega_R = 2$ moving in the opposite directions with equal speed $|v_{LR}(t_0)| \approx 0.0995$, similar to the process shown in Fig.7. After the collision, TS form a bound a ring-shaped state, in the center of which focused a dense energy concentration. The energy integral of the interactions TS system in this case is preserved with accuracy: $\frac{\Delta En}{En} \approx 10^{-3}$. Next, a similar experiment was conducted at the increased TS speed.

In Fig.9 is a description of process of head-on collision of the TS (2) of model (1): $\omega_L = 1 \rightarrow \leftarrow \omega_R = 2$ (Fig.9a); time simulation $t \in [0.0, 60.0]$; initial speed ($t_0 = 0.0$) of TS $- |v_{LR}(t_0)| \approx 0.196$. At $t = 21.0$ the interacting TS (Fig.9ab), form a bound state (Fig.9c) for a time $t \in (17.0, 27.0)$. Next, at $t \approx 30.0$ is observed a process of decay TS with $\omega_L = 1$ onto the 3 LP with topological charges $Q_t = 1$ (Fig.9d). These LP and the TS with $\omega_R = 2$ are preserved stability and moves in opposite directions from the resonance zone (Fig.9e). The energy integral of the interactions TS system in this case is preserved with accuracy: $\frac{\Delta En}{En} \approx 10^{-3}$ (Fig.9f).

Note that in our previous works (in the case of $\omega = 1$), at the speeds of interacting TS equal $|v_{12}(t_0)| \approx 0.196$ are observed decay processes of a both TS onto the LP [5-8]. But in this case, TS with $\omega_R = 2.0$, is more stable and preserved its structural integrity in the interaction.

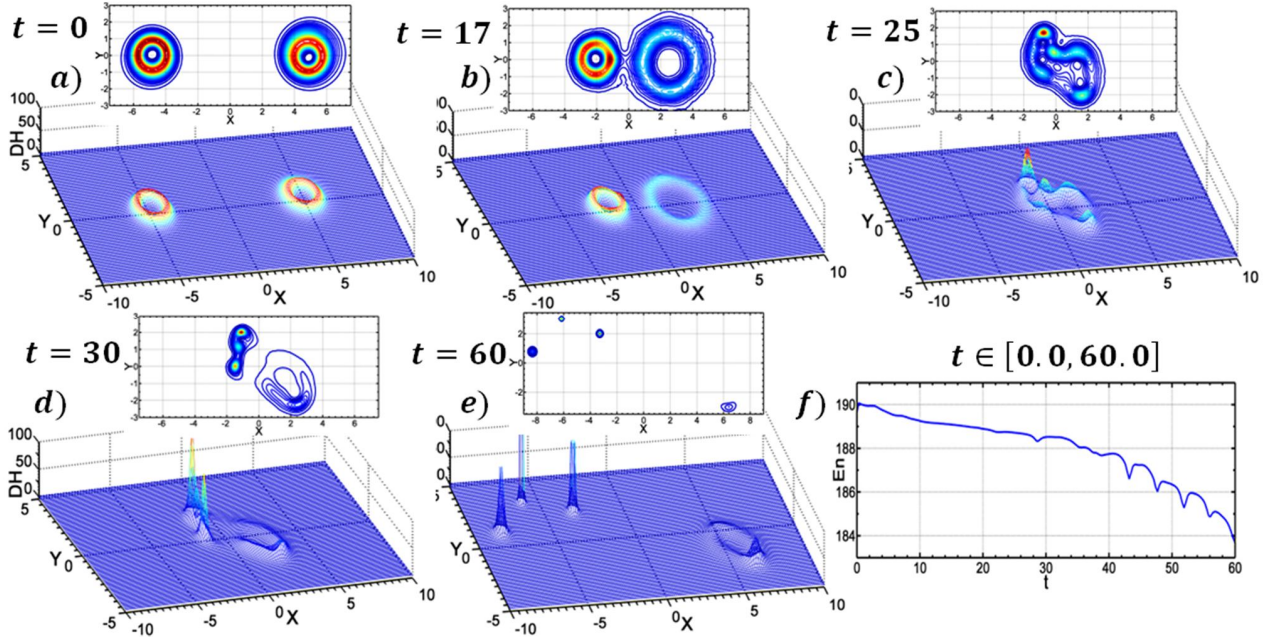

**Fig.9.** Evolution ($DH$ and its contour projection) of head-on collision of TS (2) of model (1) at $Q_{LR} = 3$, $|\vec{v}_{LR}(t_0)| \approx 0.196$, $\omega_L = 1.0$, $\omega_R = 2.0$: **a)** $t = 0.0$; **b)** $t = 17.1$; **c)** $t = 25.5$; **d)** $t = 30.6$; **e)** $t = 60.0$; **f)** energy integral for $t \in [0.0, 60.0]$.

**Case:** $\boldsymbol{\omega_L = \frac{1}{2} \rightarrow \leftarrow \omega_R = \frac{1}{2}}$. In Fig.10 shown the evolution of the TS head-on collision at $\omega_{LR} = \frac{1}{2}$; time simulation $t \in [0.0, 90.0]$; initial TS speed – $|v_{12}(t_0)| \approx 0.0995$. In this case are both TS is oscillating and moving in the opposite directions (Fig.10ab). At $t = 35.8$ are both TS moves a distance equal to about $s \approx 3.2$ units (Fig.10b) and the average loss of speed in a given period is $v_{loss}(t = 35.8) \approx 10.7\%$. Next, in Fig.10cde shows the collision and reflection process of TS where is observed (typical for this TS) some deformation of annular concentration their energy density. Because of the chirality of vortex field (2), after reflection, the directions of TS movement are changed symmetrical at a certain angle ($\alpha > 0$) (Fig.10e). The energy integral of interactions TS system in this case is preserved with accuracy: $\frac{\Delta En}{En} \approx 10^{-4}$ (Fig.10f). Next, consider the decay process of these interacting TS onto the LP.

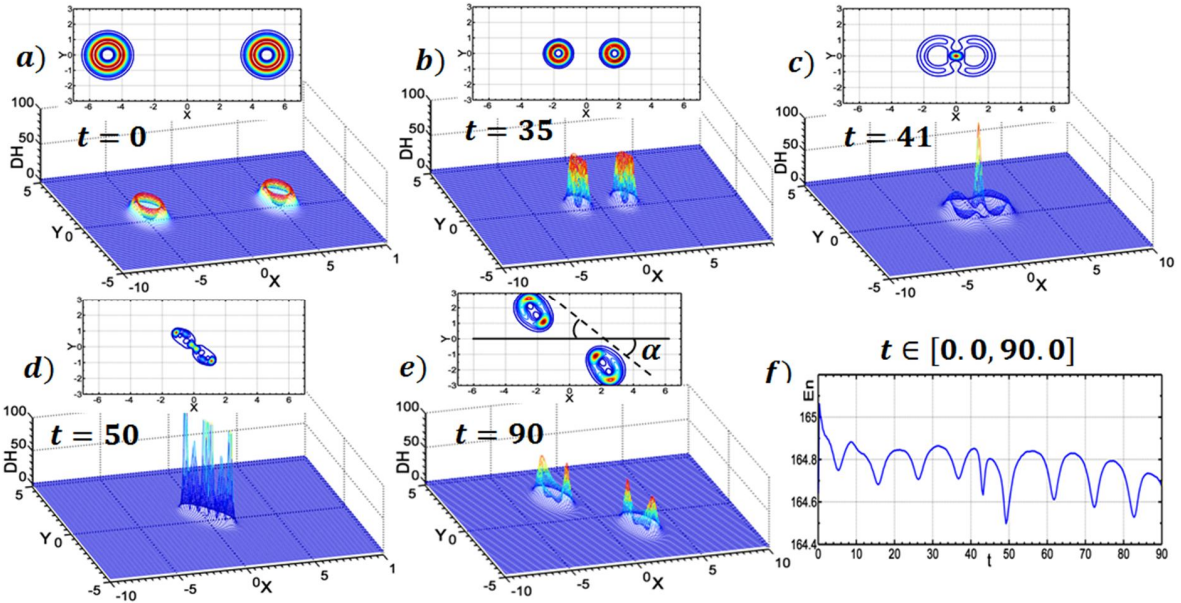

**Fig.10.** Evolution ($DH$ and its contour projection) of head-on collision of TS (2) of model (1) at $Q_{LR} = 3$, $|\vec{v}_{LR}(t_0)| \approx 0.0995$, $\omega_{LR} = 0.5$: **a)** $t = 0.0$; **b)** $t = 35.8$; **c)** $t = 41.8$; **d)** $t = 50.1$; **e)** $t = 90.0$; **f)** energy integral for $t \in [0.0, 90.0]$.

In Fig.11 shows a process similar to the previous experiment, the evolution of the TS head-on collision at $\omega_{LR} = \frac{1}{2}$; time simulation $t \in [0.0, 60.0]$. In this case, the initial speed (at $t_0 = 0.0$) of TS are increased almost twice: $|v_{LR}(t_0)| \approx$ 0.196. At $t = 16.5$ are both TS moves distance equal to about $s \approx 3.0$ units (Fig.11b) and the average loss of speed in a this time period is $v_{loss}(t = 16.5) \approx 7.24\%$. In Fig.11cde shown a process of TS collision and their decay onto the 3 pairs LP, each of the (6 LP) which has a single Hopf index: $Q_t = 1$. Note that LP does not oscillate and keep stability for quite some time: $t \in [30.0, 60.0]$.

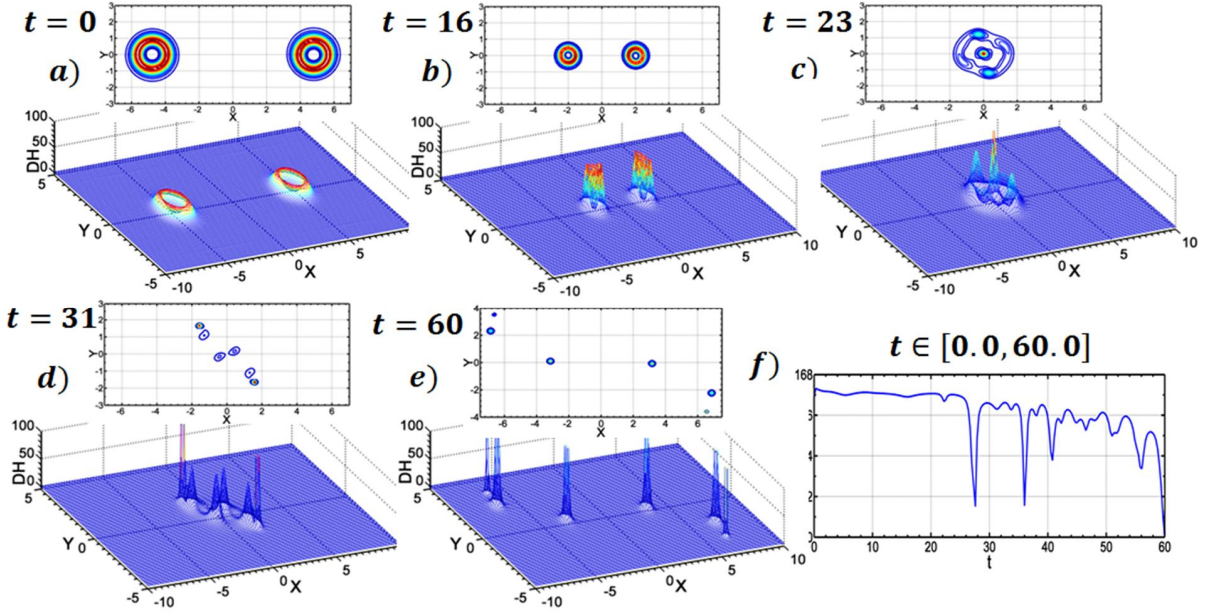

**Fig.11.** Evolution ($DH$ and its contour projection) of head-on collision of TS (2) of model (1) at $Q_{LR} = 3$, $|\vec{v}_{LR}(t_0)| \approx 0.196$, $\omega_{LR} = 0.5$: **a)** $t = 0.0$; **b)** $t = 16.5$; **c)** $t = 23.7$; **d)** $t = 31.2$; **e)** $t = 60.0$; **f)** energy integral for $t \in [0.0, 60.0]$.

**Case: $\boldsymbol{\omega_L = 2 \rightarrow\leftarrow \omega_R = 2}$.** The process of head-on interaction TS of form (2), moving with equal speed $|v_{LR}(t_0)| \approx 0.0995$ at $\omega_{LR} = 2$, similar to the process shown in Fig.10. After the collision TS are reflected from each other, and due to the presence of chirality of model are continuing the oscillation and moves in directed from the resonance zone by deflected on a certain angle trajectory. Here also is observed the deformation of annular concentration their energy density.

In the process shown in Fig.11 (at $\omega_{LR} = \frac{1}{2}$), the increase in the speed of the interacting TS until $|v_{LR}(t_0)| \approx 0.196$ leads to TS decay onto the LP (Fig.11de). But in the case $\omega_{LR} = 2$ at a similar speed increase was obtained bound

oscillating two-soliton state which remains stable in a long time: $t = 120.0$ (Fig.12).

In Fig.12a shown a process of rapprochement TS moving with equal speed in the opposite directions (head-on collision). At $t = 17.1$ are both TS moves distance equal to about $s \approx 2.5$ units (Fig.12b). The average loss of TS speed in a given time period is significant and is $v_{loss}(t = 17.1) \approx 25.4\%$. At the collision the TS forms a bound a ring-shaped state, in the center which is focused a dense energy concentration (Fig.12c).

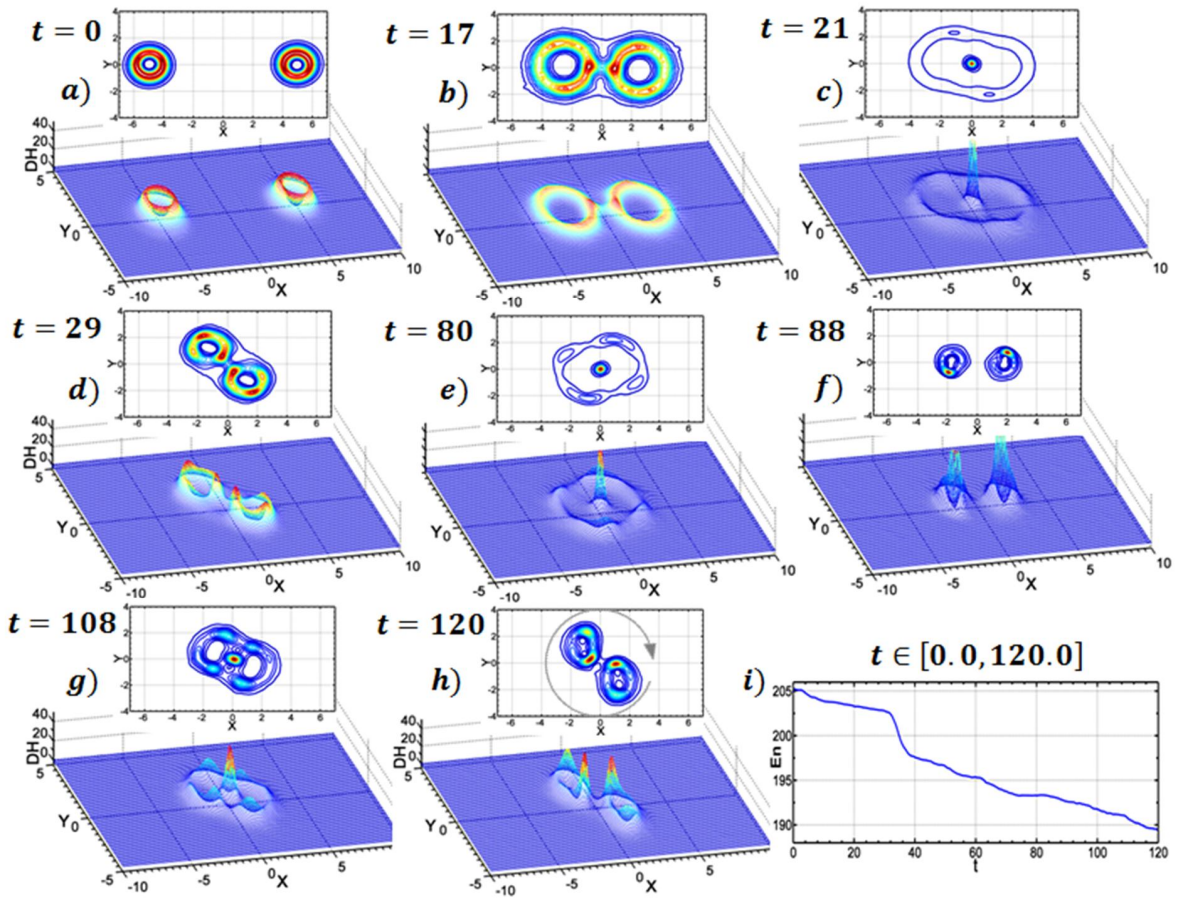

**Рис.12.** Evolution ($DH$ and its contour projection) of head-on collision of TS (2) of model (1) at $Q_{LR} = 3$, $|\vec{v}_{LR}(t_0)| \approx 0.196$, $\omega_{LR} = 2.0$: **a)** $t = 0.0$; **b)** $t = 17.1$; **c)** $t = 21.0$; **d)** $t = 29.7$; **e)** $t = 80.1$; **f)** $t = 88.2$; **g)** $t = 108.3$; **h)** $t = 120.0$; **i)** energy integral for $t \in [0.0, 120.0]$.

Next, the TS for a short time is separated (Fig.12d) and later are re-attracted, in form a bound a ring-shaped state of a similar to that shown in

Fig.12c. At simulation time ($t \in [0.0, 120.0]$), the process described in Fig.12bcd is repeated in regular intervals during 7 periods, two of which are shown in Fig.12ef and Fig.12gh. Because the chirality of the TS fields (2) a bound state of TS energy density ($DH$) apart from periodic merge-split interactions, also has a rotational movement (Fig.12h). The energy integral of the interactions oscillating TS system after collision ($t > 40.0$) is preserved with accuracy: $\frac{\Delta En}{En} \approx 10^{-3}$ (Fig.12i).

## IV. Conclusion

Thus, in this paper is solved the evolutionary problem for the dynamical equations (1), using the form vortices (2) as the initial data for various values of frequency of A3-field vector rotation in isospace of Bloch sphere. By methods of numerical simulation the oscillating TS of anisotropic 2D O(3) NLSM is obtained and their stability in the process of evolution is shown. The models of head-on two-soliton collisions of oscillating TS moving in the speed limit $\vec{v}_{LR}(t_0) \in (0.099, 0.196)$ are investigated.

In our previous works (at $\omega = 1$: topological solitons without oscillatory dynamics), at a head-on collisions of topological solitons, were observed the process of two types: collision-reflection and collision-decay onto the localized perturbations (in addition to the processes of long-range interactions and gradual annihilation, which were observed at the change of isospin dynamics of vortices) [5-8]. In case of $\omega = 1$ formation of the two-soliton bound states were observed only at noncentral head-on collisions of topological solitons moving on parallel trajectories. The results of present paper, in particular, shows that the formation of stable bound two-soliton states is also possible at head-on collisions of oscillating topological solitons (Fig.7, Fig.12). Next, the numerical experiments of present paper showed that the values of $\omega$ affects to the stability of investigated topological solitons (Fig.9, Fig12). In our experiments TS with $\omega = 2$ is more stable with respect to the topological solitons having the values $\omega = \frac{1}{2}$ or $\omega = 1$.

Investigate the properties of the interaction of the localized particle-like solutions are important, in particular, at study of the complete integrability of field-theoretic models [14].

## Acknowledgments

The author is grateful to Prof. Kh.Kh. Muminov for instructions, advice and valuable suggestions made during the discussion of the results.

## References

[1] T. H. R. Skyrme. A Non-Linear Field Theory. London: Proceedings of the Royal Society. Math. and Phys. Sc., Series A, (1961), v. 206, №1300, p. 127-138.

[2] A. M. Kosevich, B. A. Ivanov, A. S. Kovalev. Nonlinear magnetization waves: dynamical and topological solitons (in Russian). Kiev, Naukova Dumka, 1983, 193p.

[3] A. A. Belavin, A. M. Polyakov. Metastable states of two-dimensional isotropic ferromagnets. JETP, 1975, 22(10), p. 503-506.

[4] Kh. Kh. Muminov. Multidimensional dynamic topological solitons in the anisotropic nonlinear sigma model. Reports of Academy of Sciences of the Republic of Tajikistan. 2002, vol. XLV, №10, p. 28-36.

[5] Kh. Kh. Muminov, F. Sh. Shokirov. Numerical simulation of new types of topological and dynamical solitons in non-linear sigma-model. 5th Japan-Russia International Workshop MSSMBS'12. Book of Absracts. Joint Institute for Nuclear Research. Russia, Dubna, 2012, p. 47-48.

[6] Kh. Kh. Muminov, F. Sh. Shokirov. Interaction and decay of two-dimensional topological solitons in O(3) non-linear vector sigma-model. Reports of Academy of Sciences of the Republic of Tajikistan. 2011, v. 54, №2, p. 110-114.

[7] Kh. Kh. Muminov, F. Sh. Shokirov. Dynamics of interaction of two-dimensional topological solitons in the O(3) vectorial nonlinear sigma-model. Reports of Academy of Sciences of the Republic of Tajikistan, 2010, v. 53, №9, p. 679-684.

[8] Kh. Kh. Muminov, F. Sh. Shokirov. Numerical simulation of new types of topological and dynamical solitons in non-linear sigma-model. News of Science and Education / Publisher: Science and education LTD, Yorkshire, England, 2014, p. 69-77.

[9] Kh. Kh. Muminov, F. Sh. Shokirov. Dynamics of interaction of topological vortex with domain wall in (2+1)-dimensional non-linear sigma-model. Reports of Academy of Sciences of the Republic of Tajikistan, 2015, v.58, №4, p. 302-308.

[10] F. Sh. Shokirov. Dynamics of interaction of topological vortex with domain wall in (2+1)-dimensional non-linear sigma-model. Global Science and Innovation Science. Materials of the VI International Scientific Conference. USA, Chicago, November 18-19th, publishing office Accent Graphics communications. 2015, v. II, p.69-73.

[11] F. Sh. Shokirov. Dynamics of interaction of domain walls in (2+1)-dimensional non-linear sigma model. European science review. «East West» Association for Advanced Studies and Higher Education GmbH, Austria,Vienna, 9-10 (5) 2015, p.151-153.

[12] A. A. Samarsky. The theory of difference schemes [in Russian]. Moscow: Nauka, 1977, 657 p.

[13] A. A. Samarsky, P. N. Vabishevich, A.V. Gulin. Stability of operator-difference schemes. Differ. Equ., 35:2, 1999, p. 151-186.

[14] V. G. Mahankov. Solitons and numerical experiment. Particles & Nuclei, 1983, vol.14, №1, p. 123-180.